\documentclass[useAMS,usenatbib]{mn2e}
\usepackage{graphicx,color} 
\usepackage{amssymb} 
\usepackage{color} 
\usepackage{url}
\usepackage{amsmath}
\usepackage{pdflscape} 
\usepackage{newtxtext}
\usepackage{morefloats}	
\usepackage{longtable}	
\usepackage{setspace}
\usepackage{rotating}
\usepackage{float}
\usepackage{textcomp}

\usepackage[english]{babel}
\usepackage{graphicx}
\usepackage{adjustbox}
\usepackage[mediumspace,mediumqspace,Grey,squaren]{SIunits}
 \usepackage{multirow}
\usepackage{tabularx}
\newcommand{\araa}{{\it ARA$\&$A}}

\newcommand{\aap}{{\it A$\&$A}}

\newcommand{\aj}{{\it AJ}}
\newcommand{\apj}{{\it ApJ}}
\newcommand{\apjl}{{\it ApJL}}

\newcommand{\mnras}{{\it MNRAS}}

\newcommand{\pasp}{{\it PASP}}

\newcommand{\apjs}{{\it ApJS}}
\newcommand{\qjras}{{\it QJRAS}}

\newcommand{\MHI}{$M_{\rm HI}$}

\newcommand{\Mr}{$M_{\rm r}$}

\newcommand{\Mstar}{$M_{\star}$}

\newcommand{\Msun}{$M_\odot$}
\newcommand{\kms}{\mbox{km\,s$^{-1}$}}

\def\approxlt{\lower.2em\hbox{$\buildrel < \over \sim$}}
\def\approxgt{\lower.2em\hbox{$\buildrel > \over \sim$}}

\newcommand{\FHI}{\mbox{$F_{\rm HI}$}}
\newcommand{\Jykms}{\mbox{Jy~km~s$^{-1}$}}
\newcommand{\kmsMpc}{\mbox{km~s$^{-1}$~Mpc$^{-1}$}}

\newcommand{\masq}{\mbox{mag~arcsec$^{-2}$}}

\newcommand{\Lr}{\mbox{$L_{\rm r}$}}

\title{A study of the H{\LARGE I} and optical properties of Low Surface Brightness galaxies:~spirals, dwarfs and irregulars}
\author[Honey et al.]{ M. Honey$^{1,2}$\thanks{E-mail : mhoney@iiap.res.in},
W. van Driel$^3$\thanks{E-mail : wim.vandriel@obspm.fr},
M. Das$^1$\thanks{E-mail : mousumi@iiap.res.in},
J-M. Martin$^3$ 
\\
$^1$Indian Institute of Astrophysics, Koramanagala, Bangalore 560034, India.\\
$^2$Pondicherry University, R. Venkataraman Nagar, Kalapet, 605014 Pondicherry, India.\\
$^3$GEPI, Observatoire de Paris, PSL Universit\'e, CNRS, 5 place Jules Janssen, 92190 Meudon, France\\
}
\date{Accepted 2017 December 18. Received 2017 December 13; in original form 2017 November 10.}
\pagerange{\pageref{firstpage}--\pageref{lastpage}} \pubyear{2018}

\begin{document}
\maketitle

\label{firstpage}
\begin{abstract}

We present a study of the HI and optical properties of nearby ($z$ $\le$ 0.1) Low Surface Brightness galaxies (LSBGs). We started with a literature sample of $\sim$900 LSBGs and divided them into three morphological classes: spirals, irregulars and dwarfs. Of these, we could use $\sim$490 LSBGs to study their HI and stellar masses, colours and colour magnitude diagrams, and local environment, compare them with normal, High Surface Brightness (HSB) galaxies and determine the differences between the three morphological classes. We found that LSB and HSB galaxies span a similar range in HI and stellar masses, and have a similar \MHI/\Mstar\--\Mstar\ relationship. Among the LSBGs, as expected, the spirals have the highest average HI and stellar masses, both of about 10$^{9.8}$ \Msun. The LSGBs' ($g$--$r$) integrated colour is nearly constant as function of HI mass for all classes. In the colour magnitude diagram, the spirals are spread over the red and blue regions whereas the irregulars and dwarfs are confined to the blue region. The spirals also exhibit a steeper slope in the \MHI/\Mstar\--\Mstar\ plane. Within their local environment we confirmed that LSBGs are more isolated than HSB galaxies, and LSB spirals more isolated than irregulars and dwarfs. Kolmogorov-Smirnov statistical tests on the HI mass, stellar mass and number of neighbours indicates that the spirals are a statistically different population from the dwarfs and irregulars. This suggests that the spirals may have different formation and HI evolution than the dwarfs and irregulars.

\end{abstract}


\section{Introduction } \label{sec:intro} 
Low Surface Brightness galaxies (LSBGs) are late-type disc galaxies that span the same mass range as the High Surface Brightness (HSB) disc galaxies that define the Hubble sequence, but whose average surface brightness is 5-10 times  lower. The proper definition of an LSBG is through a measure of central disc surface brightness. Historically, an LSBG is defined as having a $B$-band central surface brightness fainter than 23.0 \masq\ \citep{ImpeyBothun97, Sprayberry.etal.1997}, which is 4$\sigma$ less than found originally in the central surface brightness distribution  of HSB galaxies by \citet{Freeman.1970}. This implies that LSBGs are either intrinsically rare, or heavily selected against.
Their stellar discs are generally embedded in extended HI gas discs (e.g., \citealt{mcgaugh.etal.1995}; \citealt{mishra.etal.2017}). 
\par

The LSB nature of these galaxies arises from their low disc stellar surface density ($\Sigma_{*}$), which is due to the larger average separation between stars than in normal, HSB discs, as well as from their low star formation rates (SFRs) \citep{O'Neil.etal.2004, wyder.etal.2007, Boissier.etal.2008,  zhong.etal.2010, schombert.etal.2013, young.etal.2015}. 
Their dust content is also low \citep{Hinz.etal.2007, Rahman.etal.2007} and their discs appear to be metal poor \citep{McGaugh.1994, kuziodenaray.etal.2004}. They are however rich in neutral hydrogen (HI) gas \citep{deblok.etal.1996, burkholder.etal.2001, O'Neil.etal.2004, du.etal.2015} and their HI mass to luminosity ratios are much higher than those of HSB galaxies \citep{O'Neil.etal.2004}. 
Their HI rotation curves indicate that their optical discs are embedded in very massive dark matter halos \citep{deblok.mcgaugh.1997}. Molecular gas, which is critical for star formation, is rare in LSBGs \citep{ oneil.etal.2000, das.etal.2006, das.etal.2010}. This suggests that the combination of low $\Sigma_{*}$ discs and dominant dark matter halos prevents disc instabilities from forming in these galaxies \citep{ghosh.jog.2014, honey.etal.2016}.

There are several ways to define LSBGs in practice. For our study we have used the LSBG catalogues of \citet{Schombert.etal.1988}, \citet{Schombert.etal.1992} and \citet{Impey.etal.1996} (see also \ref{sec:sample}). 

In the past few years there have been several surveys to detect LSBGs in optical images which were all selected on their low central surface brightness using the Sloan Digital Sky Survey (SDSS):  \citet{Zhong.etal.2008} identified 12,282 and \citet{shao.etal.2015} 1235 LSBGs, and \citet{williams.etal.2016} found 343 low-luminosity LSBGs for the Galaxy And Mass Assembly (GAMA) project. In the Virgo cluster, \citet{giallongo.etal.2015} and \citet{davies.etal.2016} found 303 new LSBGs and 11 ultra faint LSBGs respectively. Besides these, \citet{du.etal.2015} reported 1129 new LSBGs with large HI masses.

Morphologically, LSBGs can be broadly divided  into the classes of spirals, irregulars and dwarfs. The largest LSB spirals, which are often referred to as giant LSBGs (e.g., \citealt{das.2013}), are located in isolated environments, often near void walls \citep{Rosenbaum.etal.2009}. However the irregulars and dwarfs are found in both underdense and crowded environments (see Sect.~\ref{subsec:implications}). The origin of LSBGs is not fully understood, but  their overall isolated environments must play some role in their formation and evolution \citep{hoffman.silk.1992}. 

Although there have been many studies of LSBGs, there has not yet been a study that specifically compares the optical and cold gas properties of the abovementioned morphological subclasses of LSBGs (spirals, dwarfs and irregulars). Such a comparison is important as it can throw light on both the origin and the evolution of LSBGs. In this paper we address these issues. 

We examine the HI masses, gas mass fractions (which we define as the ratio of the total HI and stellar masses, \MHI/\Mstar), absolute magnitudes, stellar masses, optical colours and the colour magnitude diagrams of LSBGs in general and of their subclasses. We also compare these properties of LSBGs with those of normal HSB galaxies. 

Throughout this paper, unless indicated otherwise, we use distances $D$ based on a simple Hubble flow model, $D$ = $V$/$H_0$, where $V$ is the heliocentric radial velocity and the Hubble constant $H_0$ = 70 \kmsMpc.
The outline of our paper is as follows: in Section~\ref{sec:sample} we describe our sample selection, in Section~\ref{sec:analysis} we describe our analysis of the data and the results, and we present our conclusions in Section~\ref{sec:conclusion}.

\section{Sample selection} \label{sec:sample} 
Our initial sample consists of the total of 897 objects listed in the LSB galaxy catalogues of \citet{Schombert.etal.1988}, \citet{Schombert.etal.1992} and \citet{Impey.etal.1996}. 

All galaxies listed in these catalogues have known redshifts.
We set an upper limit of 0.1 to the redshifts, to study only LSBGs within the local universe, and removed the 19 objects with $z$~$>$~0.1. 

Although these catalogues have different selection criteria (see below), the objects contained in them clearly look like LSBGs on photographic plates and have been used in other studies. We prefer to use these older studies over more recent, but significantly less deep surveys (such as \citealt{Zhong.etal.2008} and \citealt{shao.etal.2015}) which are based on central disc surface brightness determinations using SDSS images \citep{York.etal.2000, Alam.etal.2015}, because their galaxies often do not have corresponding HI observations and usually no explicit identifications. 

\citet{Impey.etal.1996} defined their sample using a central surface brightness of greater than 22 magnitude arcsec$^{-2}$ in the $B$-band, from UK Schmidt Telescope IIIa-J plates measured in the automated plate measuring (APM) survey (see \citealt{Sprayberry.etal.1996} for further details).  \citet{Schombert.etal.1988} and \citet{Schombert.etal.1992} defined their samples by visual inspection of blue plates of the Second Palomar Observatory Sky Survey (POSS-II) \citep{Reid.etal.1991}, which has a surface brightness detection limit of 26.0 magnitude~arcsec$^{-2}$. Their catalogues mainly contains LSB objects that were not detected in the previous, less deep First Palomar Observatory Sky Survey (POSS-I) \citep{Minkowski.book} which has a detection limit of $\sim$ 25.2 magnitude~arcsec$^{-2}$ \citep{Cornell.etal.1987}. It includes previously catalogued UGC and NGC galaxies which they found to have LSB tails, discs, spiral arms or tails on the deeper POSS-II plates.

LSBGs span a wide range of morphological classes, from spirals, dwarfs to irregulars. We collected information on the centre positions, redshifts and optical morphological classification of the sample galaxies (see Table \ref{tab:sample}) using the NASA Extragalactic Database (NED)\footnote{http://ned.ipac.caltech.edu/}. We included the following NED morphological classifications in our three classes: Sa, Sb, Sc, Sd and Sm for the spirals, Im and Irr for the irregulars, and dE, dI and dm for the dwarfs. 
Though we have also listed 13 interacting or merging galaxies and 33 lenticulars, we have not used them for further analysis. 

The morphological classifications from NED are generally based on optical images from early photographic surveys. To verify these, we visually examined the entire sample using the  SDSS composite $gri$-band colour images from the final Data Release 12 (DR12) \footnote{http://www.sdss.org/dr12/}, see \citet{Alam.etal.2015}.
For the dwarfs, it is sometimes difficult to distinguish between disc systems and spheroidals through visual inspection of relatively shallow SDSS optical images. It is therefore possible that our dwarfs category also contains gas-poor spheroidal dE-type systems.
For 341 galaxies we modified the NED classification and listed our new determinations in Table~\ref{tab:sample}. 
\onecolumn
\begin{landscape}
\begin{center}
\begin{table*}
\caption{Global properties of the LSBGs in the initial sample of 897 galaxies -- {\it Note: sample page only, the full version is included after the References }}
\label{tab:sample}
\begin{tabular}{lcrlccccrrrl}
\hline
galaxy name & RA \& DEC & $D$ & morphology & class & HI flux & log(\MHI)& \Mr\ & ($g$--$r$) & log(\Mstar)& $Iso$ & \renewcommand{\thefootnote}{\alph{footnote}} Flag\footnote{Only the galaxies with ``good" flags are used for analysis}\\ 
 & (J2000.0) & (Mpc) & $[$NED$]$ &  & (Jy km/s) & (\Msun) & (mag) & (mag) & (\Msun) & & \\ 
\hline 
\multicolumn{12}{l}{1. With HI detections:}	 \\ 
LSBC F539-V01 	&00:11:29.00 +21:25:52.0 & 113.3 & Sm & spiral &2.37 & 9.86 & $-$18.52 & 0.34 &8.75 & 	1 & good \\
LSBC F608-V01 	&00:12:48.25 +14:31:31.1 & 26.1 & Sd & dwarf &3.47 & 8.75 & $-$14.71 & 0.24 &7.09 & 	10 & good \\
0012+0218 	&00:14:59.97 +02:34:48.3 & 256.3 & dI & dwarf &0.69 & 10.03 & $-$19.15 & 0.36 &9.02 & ... & good \\
0014+0210 	&00:16:41.72 +02:27:34.4 & 59.1 & Sm: & star	 &0.75 & 8.79 & $-$15.62 & 1.34 &8.91 & 	1 & not used\\	
LSBC F608-01 	&00:17:15.90 +17:31:23.0 & 14.3 & Im: & irregular &2.18 & 8.02 & $-$13.99 & 0.92 &7.71 & 	15 & not used \\
LSBC F473-V01 	&00:17:18.30 +26:51:41.0 & 127.3 & Im & irregular &1.36 & 9.72 & $-$18.57 & 0.33 &8.76 & 	1 & good \\
LSBC F608-V02 	&00:23:06.06 +15:08:25.6 & 78.8 & Sd & irregular &0.85 & 9.09 & $-$17.05 & 0.50 &8.37 & 	10 & good \\
LSBC F539-02 	&00:23:15.76 +20:15:59.1 & 81.7 & Im: & spiral &1.56 & 9.39 & $-$19.56 & 0.42 &9.27 & 	11 & good \\
LSBC F473-01 	&00:25:56.90 +23:55:24.0 & 80.5 & Im & irregular &0.95 & 9.16 & $-$17.83 & 0.40 &8.56 & 	3 & good \\
LSBC F473-V02 	&00:26:21.00 +24:38:35.0 & 51.1 & Irr & irregular &1.40 & 8.93 & $-$16.40 & 1.31 &9.18 & 	2 & good \\
0023+0044 	&00:26:24.98 +01:01:12.0 & 75.7 & Sd & spiral &1.51 & 9.31 & $-$19.04 & 0.49 &9.16 & 	3 & good \\
0025+0221 	&00:27:46.98 +02:38:23.5 & 58.1 & dI & dwarf &0.49 & 8.59 & $-$16.81 & 0.14 &7.81 & 	13 & good \\
0029+0037 	&00:31:43.27 +00:54:02.6 & 76.6 & & spiral &1.09 & 9.18 & $-$18.47 & 0.36 &8.75 & 	3 & good \\
0029+0226 	&00:31:45.35 +02:42:53.3 & 34.0 & dI & dwarf &1.11 & 8.48 & $-$14.38 & $-$0.50 &5.87 & 	2 & bad \\
LSBC F473-V03 	&00:34:46.10 +22:46:46.0 & 83.4 & Sm & spiral &1.19 & 9.29 & $-$17.66 & 0.55 &8.68 & 	1 & good \\
0049+0105 	&00:52:01.86 +01:21:42.9 & 184.0 & Sm & spiral &1.15 & 9.96 & $-$18.61 & 0.41 &8.87  & 	... & good \\	
0051-0227 	&00:54:21.40 $-$02:11:45.1 & 77.7 & Sd pec sp & spiral &2.65 & 9.58 & $-$17.69 & 0.30 &8.36 & 	1 & good \\
LSBC F682-V01 	&00:56:41.10 +10:20:23.0 & 148.6 & Sc & spiral &1.35 & 9.85 & $-$19.58 & 0.26 &9.07 & 	1 & good \\
LSBC F682-01 	&00:57:31.89 +10:21:48.2 & 39.4 & Sm & spiral &2.53 & 8.97 & $-$16.03 & 0.24 &7.62 & 	1 & good \\
0056+0044 	&00:58:55.46 +01:00:17.7 & 76.5 & Irr & irregular &3.38 & 9.67 & $-$17.90 & 0.13 &8.22 & 	7 & good \\
0056+0020 	&00:58:55.89 +00:36:27.9 & 77.6 & dI & dwarf &2.38 & 9.53 & $-$17.04 & 0.14 &7.90 & 	7 & good \\
LSBC F682-V02 	&01:01:51.57 +10:16:21.8 & 163.2 & Sm & spiral &0.75 & 9.67 & $-$19.71 & 0.43 &9.35 & ... & good \\
0103+0030 	&01:06:07.17 +00:46:33.7 & 74.3 & Sd & spiral &1.40 & 9.26 & $-$17.99 & 0.38 &8.59 & 	5 & good \\
0108+0242 	&01:11:21.26 +02:58:22.8 & 143.3 & Sc & spiral &1.30 & 9.80 & $-$19.98 & 0.32 &9.30 & 	1 & good \\
0110+0046 	&01:12:50.69 +01:02:48.8 & 15.8 & dI & dwarf &6.23 & 8.56 & $-$13.65 & $-$0.10 &6.10 & 	9 & not used \\
\hline
\end{tabular}
\end{table*}
\end{center}
\end{landscape}
\twocolumn

We excluded a total of 74 galaxies from further analysis, for the following reasons; they are marked accordingly in Table~\ref{tab:sample} where all are also flagged as `not used': 
(i) 44 are not visible in the SDSS images (they are marked as `no SDSS galaxy'). As these are very LSB systems, judging from the LSBGs source catalogues' data and our inspection of POSS-II plates, it is not surprising that they are not visible on the less deep SDSS images. 
Since we cannot determine their morphological types, we cannot use them for further analysis,
 (ii) three were found be star-like, (iii) two appeared too HSB (marked as `too bright'), (iv) eight were significantly displaced from their catalogued positions (marked with `location is off'), and (v) we are not sure about the morphological class for another 17 galaxies (marked as `...'). 

Our sample of LSB galaxies with morphological classification thus contains 758 galaxies, of which 426 (56\%) are spirals, 107 (14\%) are irregulars and 225 (30\%) are dwarfs. Hereafter this sample is referred to as Sample~I.

\section{Analysis and results} \label{sec:analysis}
The following global properties of the initial sample of 897 galaxies are listed in Table~\ref{tab:sample}, see the text for further details: 

\begin{itemize}
\item{galaxy name}: catalogue name taken from the LSBGs samples from which the galaxy was selected (see Sect.~\ref{sec:sample}); 
\item{RA \& DEC}: Right Ascension and Declination of the SDSS spectroscopic source in epoch J2000.0 coordinates;
\item{$D$}: distance (in Mpc) = $V/H_{\mathrm 0}$, where $V$ is the heliocentric radial velocity in \kms, for a Hubble constant $ H_{\mathrm 0}$= 70 \kmsMpc;
\item{morphology $[$NED$]$}: optical morphological classification, taken from NED;
\item{morphological class}: global optical morphological class (spiral, irregular, dwarf), based on our visual inspection of SDSS DR12 images;
\item{HI flux}: \FHI, integrated HI line flux (in \Jykms); 
\item{log(\MHI)}: total HI mass (in \Msun); 
\item{\Mr}: absolute $r$-band magnitude corrected for Galactic extinction following \citet{Schlegel.etal.1998}, as also used by SDSSDR12); 
\item{($g$--$r$)}: integrated ($g$--$r$) colour corrected for Galactic extinction following \citet{Schlegel.etal.1998}, as also used by SDSS DR12;
\item{\Mstar}: total stellar mass, estimated following \citet{Bell.etal.2001} (in \Msun);
\item{$Iso$}: number of near neighbours within 1 Mpc radius and $\pm$500 \kms\ from the systemic velocity of the target galaxy;
\item{Flag}: (i) on SDSS photometry: `good' indicates no BADSKY, BRIGHT, NOPROFILE, SATURATED, SUSPICIOUS DETECTIONS (like cosmic ray or CCD artifact-MAYBE-CR, MAYBE-EGHOST), or INTERPOLATION ERRORS (PSF-FLUX-INTERP) flags; `bad' indicates the target has (at least) one of these flags, or (ii) `not used' indicates either not included in Sample~I (see Section \ref{sec:sample}) or fainter than the absolute magnitude limit \Mr\ = $-$14 mag (see Section \ref{subsubsec:Re-examination}).
\end{itemize}

\subsection[]{H\,{\sevensize\bf I} mass} \label{subsec:HImass} 
The integrated, mean homogenized 21cm HI line fluxes (\FHI, in Jy \kms) were taken from the online HyperLeda database\footnote{http://leda.univ-lyon1.fr/} (based on their $m_{\rm 21}$ values,  see also \citealt{Paturel.etal1.2003} ; \citealt{Paturel.etal2.2003}). 
We found that 449 galaxies from Sample~I have published HI detections. Of these, 246 are classified as spirals, 122 as dwarfs and 81 as irregulars. The remaining 309 galaxies do not have HI data listed in HyperLeda. As we cannot determine from the HyperLeda data if they are either not observed or not detected, we have marked them all as ``without HI data" in our plots and tables. 

\begin{figure*} 
\centering
\includegraphics{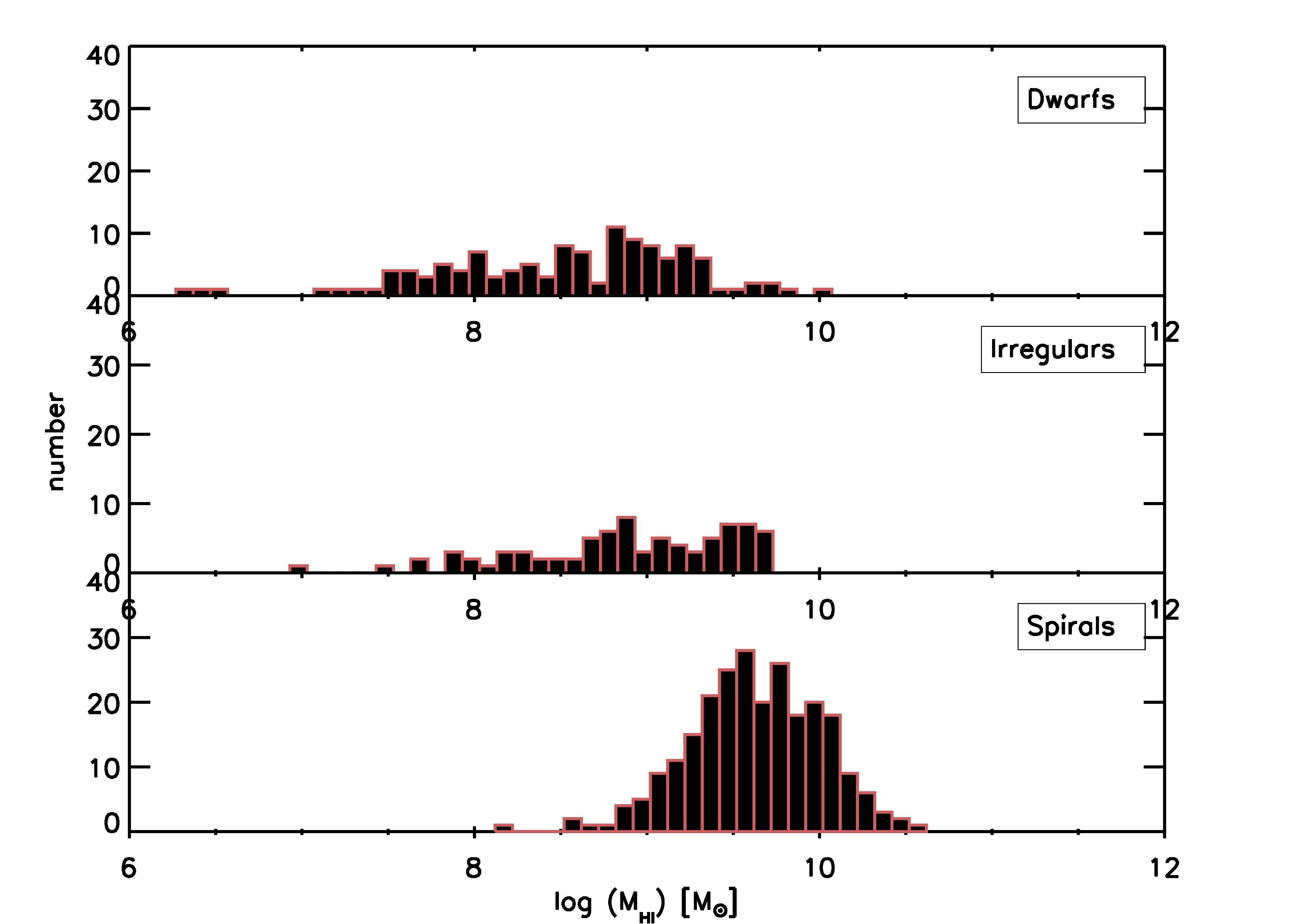}
\caption{Histograms of the total HI mass distributions (log(\MHI), in \Msun) of the three morphological classes into which we divided our LSB galaxy sample: dwarfs, irregulars and spirals. We used 246 spirals, 81 irregulars and 122 dwarfs for this plot. The bin size in log(\MHI) = 0.10.}
\label{fig:HIhistogram}
\end{figure*}

We determined the total HI masses, \MHI, using the relationship \MHI\ = 2.36$\times$10$^5$ D$^2$ \FHI\ (\Msun), where $D$ is the distance in Mpc and \FHI\ the integrated HI line flux (in \Jykms).
 
\subsubsection{H\,{\sevensize\it I} mass distributions} \label{subsubsec:HImass distribution}
LSBGs are HI rich in nature, and they have on average an $\sim$8 times higher \MHI/$L$ ratio than HSB galaxies, sometimes exceeding 10 \Msun/{$L_{\odot,{\rm B}}$} \citep{O'Neil.etal.2004}. 
The HI masses of our sample span a wide range, from 10$^7$ to 10$^{11}$ \Msun\ (see Fig. \ref{fig:HIhistogram}), 
with the spirals at the high end, the dwarfs at the low end and the irregulars in between. The mean HI masses and their standard deviations of the spirals, irregulars and dwarfs are log(\MHI/\Msun)~= 
9.78$\pm$0.40, 9.20$\pm$0.42 and 8.95$\pm$0.69, respectively. 
Thus LSBGs clearly have massive HI reservoirs, which may be linked to their low star formation rates \citep{Boissier.etal.2008, wyder.etal.2009}. 

\subsection{Absolute magnitudes and colours} \label{subsec:magnitudes} 
The basic photometric data of our sample galaxies are taken from the SDSS DR12 archival data. We retrieved apparent model magnitudes in the $g$, $r$ and $i$-bands together with the Galactic extinction coefficients from \citet{Schlegel.etal.1998} for these bands. 

\subsubsection{Galaxies with underestimated SDSS magnitudes} \label{subsub:underestimated magnitudes}
Low Surface Brightness galaxies, by their very nature, are faint. Hence, it is quite easy and natural for most imaging studies to miss extended, very LSB flux in these objects at levels where it effectively competes with the noise in the night sky. This can have serious consequences for determining their total magnitude, as in some cases (like for large disc scale lengths) the bulk of the luminosity comes from outer regions fainter than 25 \masq.

The SDSS is no exception in this respect. Based on our experience with the NIBLES HI survey of total stellar-mass selected SDSS galaxies \citep{van2016}, we looked for objects for which the total SDSS magnitudes appear to be ``misunderestimated", which can lead to (severely) reduced luminosities and hence unphysically high \MHI/$L$ ratios.

For galaxies that are extremely faint we know there is a high probability that their parameters are erroneous. For this reason we only included galaxies in our sample that are more luminous than \Mr\ = $-$14 (after re-examination of their photometry as described hereafter).. 

For 290 galaxies in sample~I we found that their listed absolute magnitudes were fainter than $-$14, which appears too low for the kind of objects we are studying, and that their \MHI/\Lr\ ratios were unphysically high, up to over a thousand.
As in the case of NIBLES, we found that among these, 130 are dwarf-like objects of small optical apparent size  that have numerous SDSS photometric sources within their outlines (see Fig.~\ref{fig:faintgalaxies}). 
Some of these  sources have Petrosian radii much smaller than the apparent size of the galaxy, indicating that the magnitudes of these sub-structure sources cannot represent the total brightness of the entire galaxy.
Also like in NIBLES, we found 160 luminous galaxies where the SDSS spectroscopic and photometric source was in fact a distinct sub-structure outside the central region, which also leads to ``misunderestimated" total magnitudes.
We therefore re-examined all these 290 galaxies in order to check whether their magnitudes are underestimated or if they are really intrinsically faint. 

\subsubsection{Re-examination of galaxies with underestimated magnitudes } \label{subsubsec:Re-examination}
First of all, the $\sim$ 290 galaxies that have an absolute $r$-band magnitude \Mr\ $>$ $-$14 were extracted from the original sample. They were all re-examined using their composite SDSS~DR12 $rgb$ colour images.
For each galaxy we found there is more than one photometric source associated with it, sometimes of quite different magnitudes. We applied the same method used for NIBLES to assess and correct the ``misunderestimated" magnitude problem. For each galaxy, we searched for all photometric sources whose centre positions lie within 30$^{\prime\prime}$ distance from the galaxy's centre coordinates, 
and selected the brightest. We again calculated the total absolute magnitude, now using only the brightest source per galaxy. We also used this method to correct the total magnitude of the 271 more luminous galaxies whose underestimated ``total" magnitudes were those of sub-structures within their discs.  

\begin{figure*} 
\centering
\includegraphics[scale=0.7]{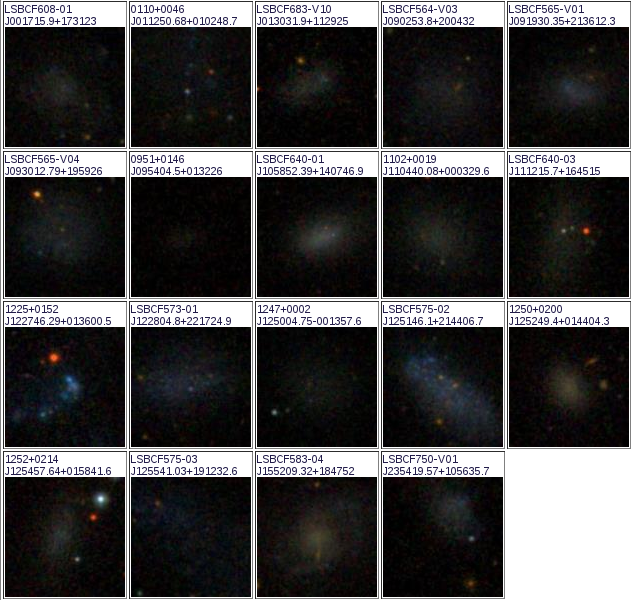} 
\caption{SDSS composite $gri$ colour images of the 19 LSB galaxies with ``misunderestimated" luminosities (see Sect.~\ref{subsub:underestimated magnitudes}), whose corrected $r$-band absolute magnitude based on the brightest SDSS photometric source within its outlines still remains fainter than $-$14. The size of each image is 45$''$ $\times$ 45$''$. In all images north is up and east is left. The galaxy catalogue names as used in the tables and their SDSS coordinates are shown in the top-left corner of each image.}
\label{fig:faintgalaxies}
\end{figure*}

For only 15\% (19/130) of the dwarf-like galaxies we found that their absolute magnitudes using the brightest source were still fainter than $-$14. This may be due to an error in the magnitude measurement, or an intrinsic property of the galaxy. Images of these 19 low-luminosity galaxies are shown in Fig.~\ref{fig:faintgalaxies} and their properties are listed in Table~\ref{tab:faint_galaxies}. Since these galaxies are fainter than our lower magnitude limit, we did  not use these galaxies for further optical studies.

The criteria for the ``good" SDSS photometry flag (see Sect.\ref{sec:analysis}) removed a further 117 galaxies from our previously defined Sample~I , leaving a total of 622 galaxies to be used for further analysis, referred to as Sample~II hereafter.

\begin{table}  
\caption{All galaxies with corrected \Mr\ fainter than  $-$14 mag }
\label{tab:faint_galaxies}
\begin{tabular}{lcc}
\hline
galaxy name & \Mr\ & Petrosian radius \\
 	    & (mag)& ($\arcsec$)   \\
\hline
LSBC F608-01  & $-$13.99	& 17.3	 	\\ 
0110+0046     & $-$13.64	& 17.3	 	\\ 	
LSBC F683-V10 & $-$13.82	&  8.7	 	\\  
LSBC F564-V03 & $-$12.63	& 43.8		\\ 
LSBC F565-V01 & $-$12.69 	&  1.2	 	\\ 	
LSBC F565-V04 & $-$13.09	& 1.2	 	\\ 
0951+0146     & $-$12.71	&  1.2 	 	\\ 
LSBC F640-01  & $-$13.48	&  6.7 		\\ 
1102+0019     & $-$13.65  	& 17.3	 	\\ 
LSBC F640-03  & $-$13.70	&  1.2	 	\\ 
1225+0152     & $-$12.71  	&  1.9	 	\\ 
LSBC F573-01  & $-$13.27 	& 17.3 	 	\\ 
1247+0002     & $-$13.07	& 17.3		\\ 
LSBC F575-02  & $-$13.96	& 10.3	 	\\ 
1250+0200     & $-$13.50 	&  5.6	 	\\ 
1252+0214     & $-$13.86	& 17.8 		\\ 
LSBC F575-03  & $-$12.46	& 27.0 		\\ 	
LSBC F583-04  & $-$13.89	&  0.9 	 	\\  	
LSBC F750-V01 & $-$13.92	&  6.7	 	\\ 	
\hline
\end{tabular}
\end{table}

\subsubsection{$r$-band luminosity as a function of H\,{\sevensize\it I} mass} \label{subsubsec:mr_vs_HImass}
\begin{figure} 
\centering
\includegraphics[scale=0.5]{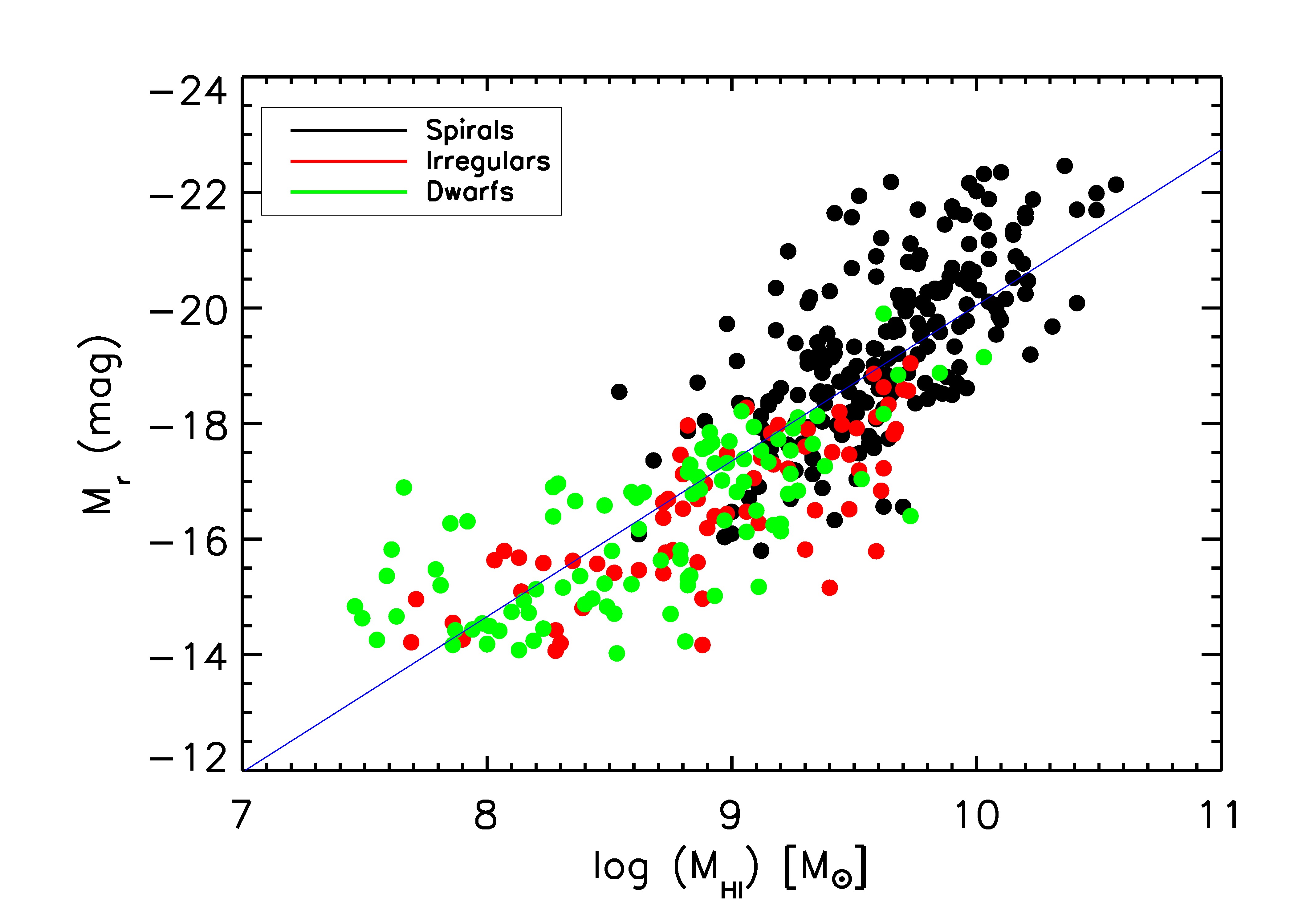}
\caption{Absolute $r$-band magnitude corrected for Galactic extinction as a function of total HI mass (log(\MHI), in \Msun) for the three morphological classes of LSB galaxies: spirals (black dots), irregulars (red dots) and dwarfs (green dots). The blue line shows the linear LSQ fit to all data. We used 206 spirals, 71 irregulars and 98 dwarfs for this plot.}
\label{fig:Mr_vs_HI}
\end{figure}

The absolute $r$-band magnitude, \Mr, as a function of total HI mass is shown in Fig.~\ref{fig:Mr_vs_HI} for LSB spirals, dwarfs and irregulars. All morphological types show a similar correlation, with more luminous galaxies having more gas; a similar relationship is observed in the $B$-band \citep{Sprayberry.etal.1995, Neil.etal.2000}.

A linear LSQ fit to all data (blue line in Fig.~\ref{fig:Mr_vs_HI} ) has a slope of $-$2.69$\pm$0.09 with an intercept of 6.89$\pm$0.88.

\subsubsection{Colour as a function of H\,{\sevensize\it I} mass} \label{subsubsec:colour_vs_HImass} 
\begin{figure} 
\centering
\includegraphics[scale=0.5]{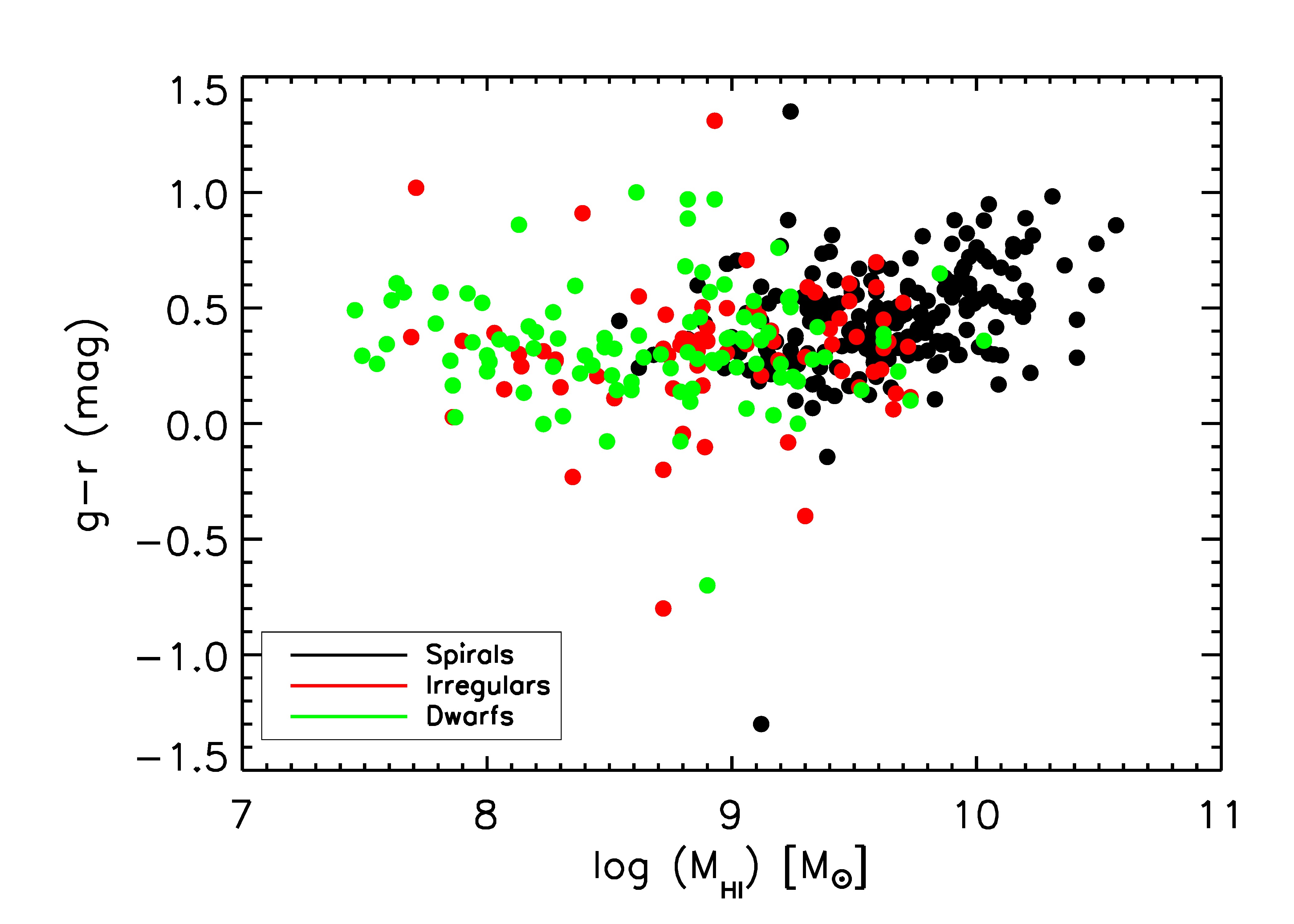}
\caption{Integrated ($g$--$r$) galaxy colour corrected for Galactic extinction as a function of total HI mass (log(\MHI), in \Msun) for the three morphological classes of LSB galaxies: spirals (black dots), irregulars (red dots) and dwarfs (green dots). The number of galaxies used in this plot is similar to Fig.~\ref{fig:Mr_vs_HI}}
\label{fig:colour_vs_HI}
\end{figure}

We examined how the integrated ($g$--$r$) galaxy colours vary as a function of total HI mass (see Fig.~\ref{fig:colour_vs_HI}) but did not note any obvious global correlation for dwarfs and irregulars. Our result is similar to earlier studies using ($B$--$V$) colours \citep[e.g.,][]{Neil.etal.2000}. The reason for this could be the low star formation rate in these galaxies: even though they contain large quantities of HI, the combined stellar and gas surface density is still lower than the threshold value for star formation \citep{das.etal.2010}. In earlier studies (e.g., \citealt{mishra.etal.2015}; \citealt{Schombert.McGaugh.2015}), it is also found that some LSB discs undergo episodic star formation, lasting for only  $\sim$10$^8$ years, which makes the galaxies bluer due to the sudden injection of young, massive stars.

However, the data suggest that the HI-rich LSB spirals (with log(\MHI) $>$ 9.8) have ($g$--$r$) colours that are on average approximately 0.5 magnitude redder compared to the lower-luminosity spirals, dwarfs and irregulars.

It has been noted previously \citep{Sprayberry.etal.1995} that the ($B$--$V$) colours of giant LSBGs are redder than those of average-sized LSB galaxies, but similar to HSB galaxies. Earlier studies have also shown the existence of red galaxies that are HI-rich, both LSBGs  \citep{vanderHulst1987, O'Neil1997, Neil.etal.2000} and HSB spirals (e.g., \citealt{Schommer.1983}; \citealt{van2016}). 

The stellar content of giant LSB spirals is more dominated by old stars compared to the LSB dwarfs and irregulars \citep{Sprayberry.etal.1995}. It may also be due to the presence of bright bulges that are seen in giant LSB spirals like UGC 6614 and Malin 2.

\subsection{Total stellar masses} \label{subsec:stellarmass} 

We compared the distributions of the total stellar mass, \Mstar, and the HI to the stellar mass ratio, \MHI/\Mstar, of all three morphological classes, since it can also tell us about their ability to convert gaseous fuel into stars. 

We converted the SDSS ($g$--$r$) colours into Johnson-Kron-Cousins system ($B$--$V$) colours following the standard conversion relation given by \citet{Jester.etal.2005}. Using the $r$~band luminosities and ($B$--$V$) colours, we estimated total stellar masses using the $r$-band \Mstar/\Lr\ ratio given in \citet{Bell.etal.2001}. 
 Since the galaxies are non-interacting in nature, we used the closed box model for this calculations. The stellar masses obtained are listed in Table~\ref{tab:sample}. 

\begin{figure*} 
\centering
\includegraphics[scale=0.8]{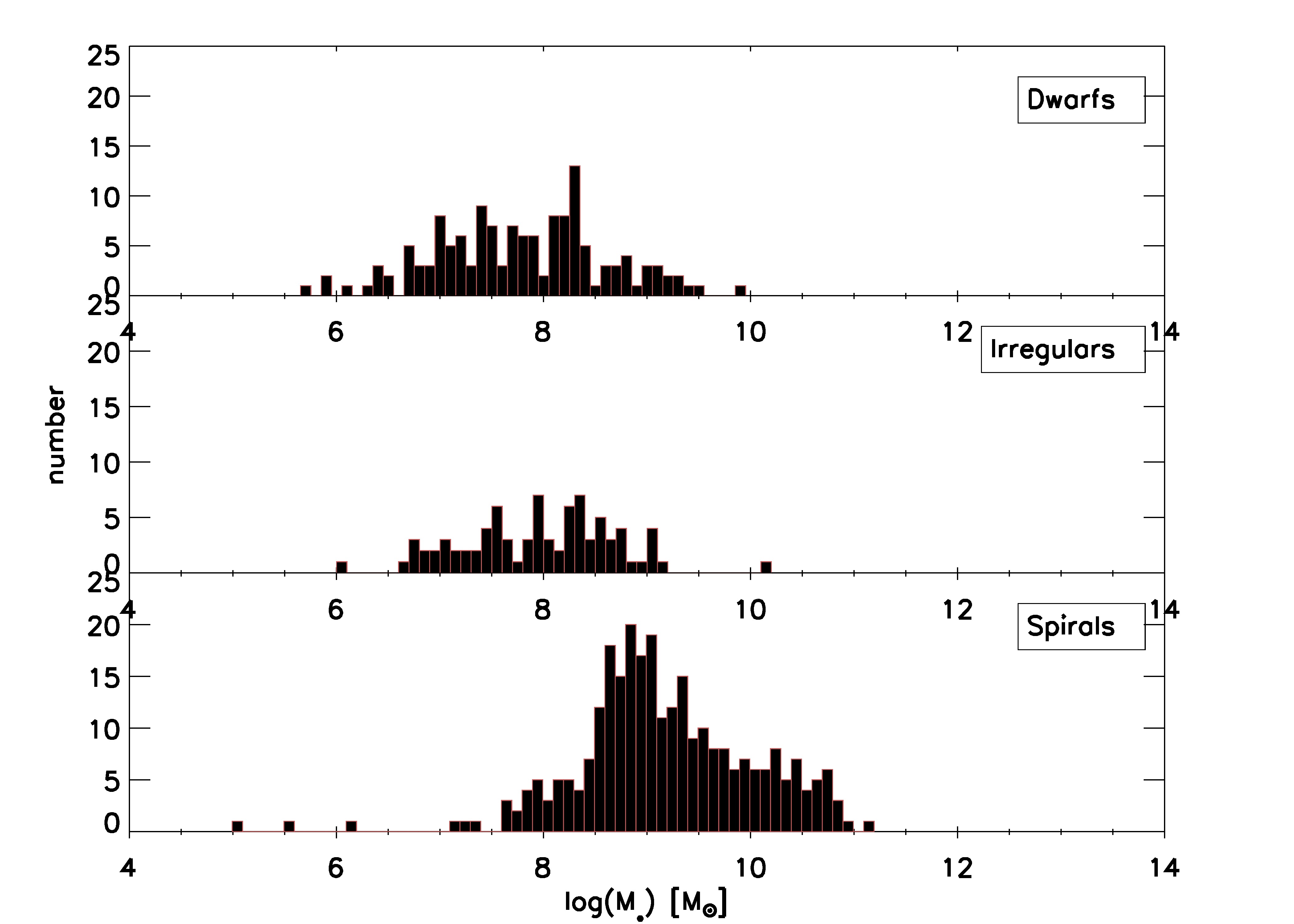}
\caption{Histograms of the total stellar mass distributions (log(\Mstar), in \Msun)  of the three morphological classes of LSB galaxies: dwarfs, irregulars and spirals. We used 283 spirals, 83 irregulars and 142 dwarfs for this plot. The bin size in log(\Mstar) = 0.10.} 
\label{fig:stellar_histogram}
\end{figure*}
Histograms of the total stellar mass distributions are given in Figure~\ref{fig:stellar_histogram}. The \Mstar\ distribution of spirals spans a wide range, 10$^6$ to 10$^{12}$ \Msun, with a mean value and standard deviation of log(\Mstar/\Msun) ~=~9.87$\pm$0.92. For irregulars and dwarfs their mean values are similar, 8.59$\pm$1.70 and 8.47$\pm$1.15 respectively, and on average about 20 to 25 times lower than for spirals.

\subsection{Colour Magnitude Diagram} \label{subsec:CMD} 
For the colour magnitude diagram (CMD), a useful tool to assess the evolutionary stage of a galaxy, we plotted the integrated ($g$--$r$) colours of galaxies as a function of \Mr\ (see Fig.~\ref{fig:CMD}). The CMD divides SDSS galaxies in the local Universe into two main regions, the blue cloud and the red sequence (e.g., \citealt{baldry.etal.2004}; \citealt{wyder.etal.2007}). The blue cloud is composed of gas-rich galaxies, such as late-type spirals, that have ongoing star formation. Their sizes are usually smaller and their absolute red magnitudes are lower than for the red sequence galaxies. The red sequence is dominated by early-type (elliptical) galaxies, which usually have little or no gas, no ongoing star formation, and are dominated by an old stellar population. To compare the CMDs of LSB and HSB galaxies in the local Universe, we first constructed a CMD of 52,000 SDSS galaxies in the local Universe, which is dominated by HSB objects. We selected them on criteria similar to those of our LSBGs: \Mr\ $<$ $-$14 and $z$ $\le$ 0.1.
We then superimposed all 491 LSBGs from our Sample II with known SDSS colours and absolute magnitudes (375 with HI  detections and 116 without HI  data) on the distribution of the 52,000 predominantly HSB SDSS galaxies, see Figure~\ref{fig:CMD}.

\subsubsection{Distribution of LSBGs in the Colour Magnitude Diagram} \label{subsubsec:LSBGs_CMD}
\begin{figure*} 
\centering
\includegraphics[scale=0.65]{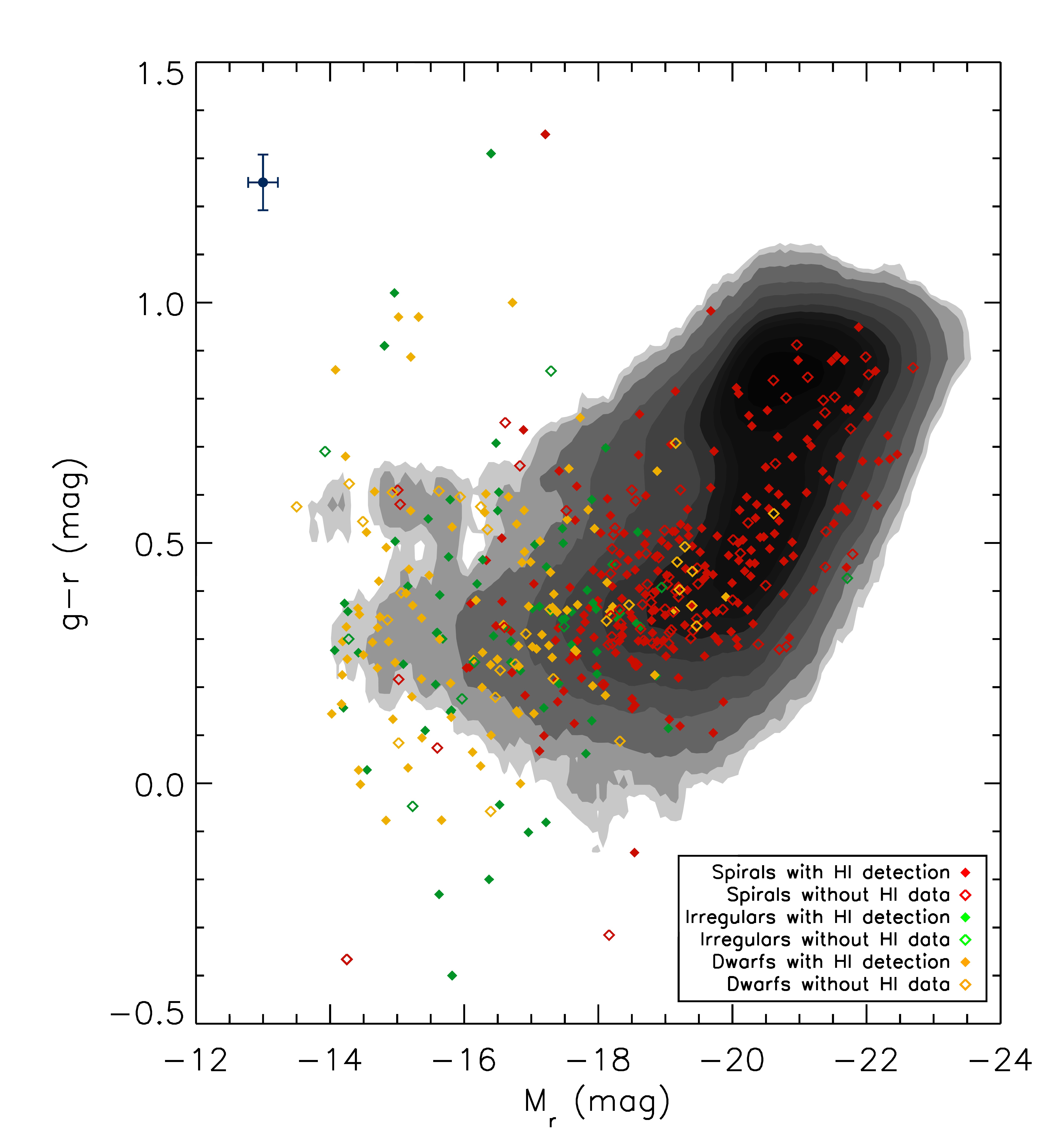}
\caption{Colour Magnitude Diagram  of all 491 LSBGs from our Sample I with known SDSS colours (375 HI detections and 116 galaxies without HI data) and of a comparison sample of 52,000 SDSS galaxies at redshifts $z<$0.1, which are predominantly HSB objects. Plotted are integrated ($g$--$r$) colours as a function of the absolute magnitudes in the $r$-band, \Mr; both properties were corrected for Galactic extinction. Our LSBGs are shown as individual coloured points superimposed on a greyscale contour map of the distribution of the large comparison sample, where the darker shades represent the denser regions; the density contour levels are in units of the number of galaxies per grid unit (of size 0.078 magnitude in \Mr\ and 0.028 magnitude in $g$--$r$): 0.05,0.1,0.2,0.5,0.7,1,2,5,10,20,30,40,50,75,100,150,200,300,400,500,600 and 700. 
The different morphological classes of LSBGs are indicated by different colours: red for spirals, green for irregulars and yellow for dwarfs. Filled symbols indicate LSBGs with HI detections, open symbols those without HI data. The typical uncertainties are represented by  the blue cross in the upper left corner. }
\label{fig:CMD}
\end{figure*}

In Fig.~\ref{fig:CMD} the red sequence lies in the upper-right part of the galaxy distribution, and the blue cloud in the lower-left.
A caveat in the interpretation of diagrams such as these, which use integrated colours, is the mixing of disc- and bulge colours. The pure disc colours can only be measured using bulge/disc decomposition on individual galaxies' luminosity profiles. Such a task is outside the scope of the present study. We would like to point out that only in case of prominent bulges will the entire galaxy appear significantly redder than its disc's colours.

The plot shows that the spiral LSBGs are distributed throughout both the blue and red regions, whereas the irregulars and dwarfs are predominantly found in the blue cloud. We note that a significant fraction of the gas-rich spirals are located in the red sequence region, which is usually dominated by gas-poor systems such as ellipticals, although gas-rich LSB and HSB spirals are also known to exist there  (see Sect. \ref{subsubsec:mr_vs_HImass}). The dwarfs and irregulars are low luminosity, gas rich systems. 

The difference in the distribution of spirals versus irregulars and dwarfs in the CMD suggests that both their stellar compositions and evolutionary histories are different. Large bulges could be the reason why some spiral LSBGs lie in the red part of the CMD: a large fraction of spiral LSBGs have massive bulges which sometimes host active galactic nuclei \citep{schombert.etal.2001, ramya.etal.2011, subramanian.etal.2016} and faint discs that show little star formation. 

\subsubsection{Gas mass fraction as a function of stellar mass} 	\label{subsubsec:HI_fraction}
\begin{figure} 
\centering
\includegraphics[scale=0.5]{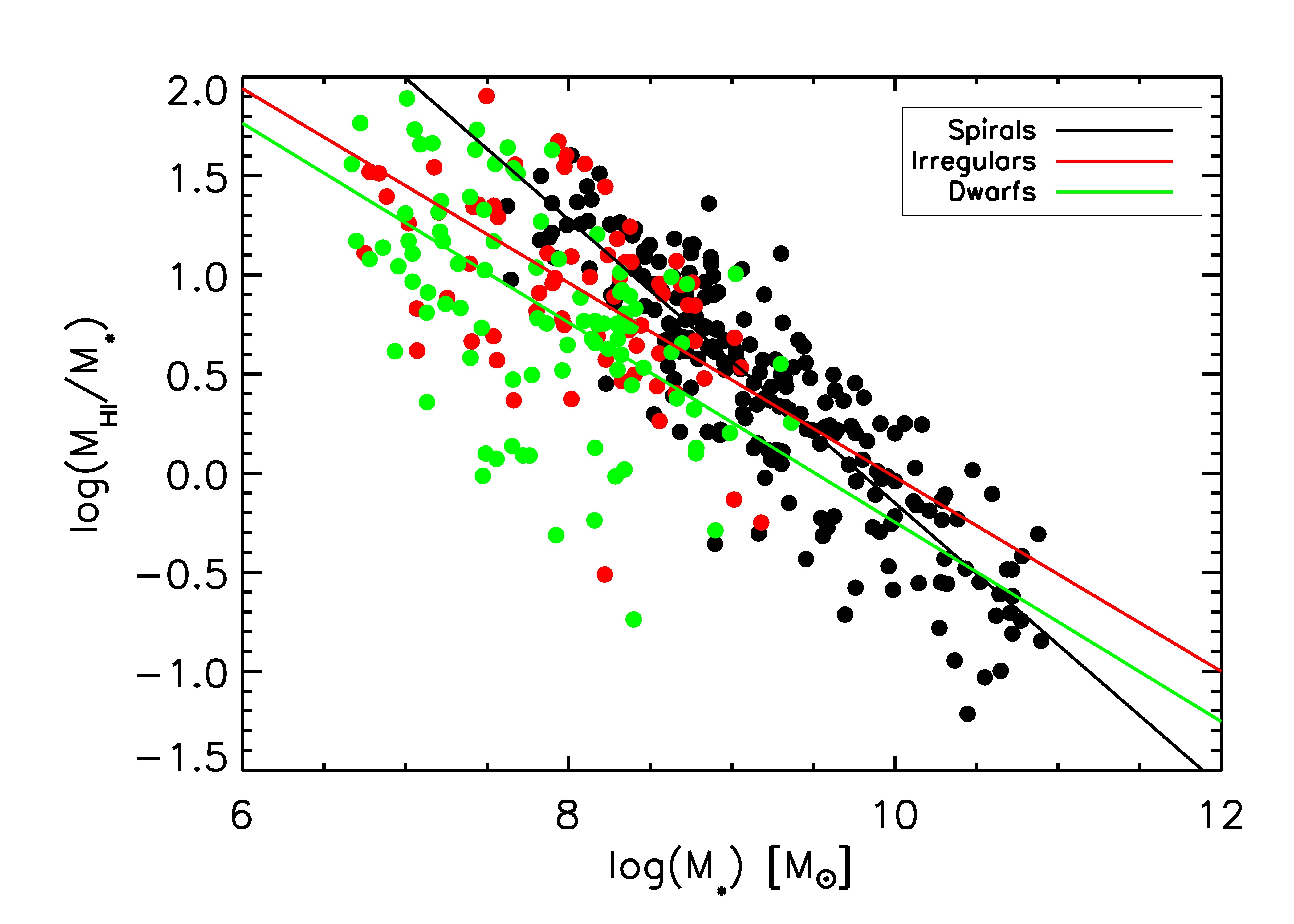}
\caption{Gas mass fraction, log(\MHI/\Mstar), as a function of total stellar mass (log(\Mstar), in \Msun). The three morphological classes of LSBGs and the linear LSQ fits to their data  are indicated by different colours: red for spirals, green for irregulars and yellow for dwarfs. For the slopes and intercepts of the three fits, see Sect.~\ref{subsubsec:HI_fraction}. The number of galaxies used in this plot is similar to Fig.~\ref{fig:Mr_vs_HI}} 
\label{fig:HI_fraction_stellar}
\end{figure}

The HI-to-stellar mass fraction, \MHI/\Mstar, is plotted as a function of the stellar mass in Figure~\ref{fig:HI_fraction_stellar}. As the stellar mass increases the fraction decreases. Of the three classes, the spirals' \Mstar\ extends to the highest values, as expected. Overall, the irregulars and dwarfs have a larger HI mass fraction than spirals and hence a larger fraction of unused neutral hydrogen. The dwarfs show the largest scatter among the three classes, in particular towards relatively gas-poor objects with low \MHI/\Mstar\ ratios.  Whereas the latter can be due to gas-poor dwarf ellipticals, very high gas fractions are also expected among dwarfs as they are one of the most dark matter dominated types of galaxies known \citep{schombert.etal.2001}, and dark matter halos can slow down their SFR by preventing the formation of disc instabilities \citep{mihos.etal.1997} which can trigger the onset of star formation. 

We compared our results with two samples of normal galaxies, one of HI detections only \citep{Papastergis.etal.2012} and another, NIBLES, selected on total stellar mass \citep{van2016}. 

It should be noted that each of the three studies use a different method to estimate \Mstar: for NIBLES the MPA/JHU catalogues \citep{Kauffmann.etal.2003, Brinchmann.etal.2004} were used, \citet{Papastergis.etal.2012} fitted all five SDSS photometric band magnitudes using model spectral energy distributions (see \citealt{Huang.etal.2012}), whereas we followed \citet{Bell.etal.2001}. In general, different methods for stellar mass estimates can differ systematically by as much as 0.3 dex whereas for individual galaxies the scatter can be up to 0.6 dex \citep{Papastergis.etal.2012, Pforr.etal.2012}.

We could  not reduce the HI flux scales of the three studies to the same reference scale, as the HyperLeda literature data used for our survey come from many different publications. The HI flux scale used by \citet{Papastergis.etal.2012} was found to be on average a factor 1.45 higher than the scale used by NIBLES, which is based on \cite{O'Neil.2004.AJ}; this factor corresponds to 0.16 in log(\MHI).

The linear regression fit to the HI detections of the NIBLES sample has a slope of $-$0.54 and an intercept of 4.70 and the fit by \citet{Papastergis.etal.2012} has a slope of $-$0.43 and an intercept of 3.75 (see \citealt{van2016}).

We did linear LSQ fits to our data points in Fig.~\ref{fig:HI_fraction_stellar}. The spirals have a slope of $-$0.71$\pm$0.03 and an intercept of 7.00$\pm$0.24, which is steeper than that of the irregulars and dwarfs. For irregulars the slope is $-$0.49$\pm$0.08 and the intercept 4.88$\pm$0.61; the values are similar for the dwarfs which have a slope of $-$0.50$\pm$0.07 and intercept of 4.79$\pm$0.57. Within the uncertainties, the fits to our LSB spirals are similar to those of the two samples which consist mainly of HSB spirals \citep{Papastergis.etal.2012, van2016}.

\subsection{Local environmental properties} \label{subsec:Environment}		 

It is well established that the large scale galaxy environment plays an important role in shaping the morphology of galaxies \citep{dressler.1980}, as well as in determining their colours \citep{bamford.etal.2009}. However, the local environment around a galaxy is also very important for its evolution. Having a larger number of near-neighbours increases the possibility of a galaxy undergoing interactions and mergers, which can result in increased star formation \citep{park.etal.2007} associated with the tidal effects and disc instabilities such as bars and spiral arms. For isolated galaxies the possibility of interactions is lower, and for these systems slow, internal evolution -- called secular evolution -- becomes more important. The giant LSBGs are often found to fall into the latter category (\citealt{das.2013}, see also Sect.~\ref{subsec:glsb}). 

Earlier studies of the local environment of LSBGs found that they are more isolated than HSB galaxies (\citealt{Schombert.etal.1988}; \citealt{Schombert.etal.1992}; \citealt{Bothun.etal.1993}; \citealt{Galaz.2011}), but more recent studies show that LSB dwarfs and irregulars are found in both high- and low density environments \citep{javanmardi.etal.2016}, whereas the larger LSB spirals are mainly located in low density regions and often close to the edges of voids \citep{Rosenbaum.etal.2009}.

We studied the local environment of each of our LSBGs by determining the number of neighbouring galaxies within 1 Mpc radius and within $\pm$500 \kms\ of its systemic velocity. We used NED radial velocity constrained cone searches\footnote{http://ned.ipac.caltech.edu/forms/denv.html} for finding neighbours, which is necessarily limited to nearby galaxies with known redshifts. This means that the number of neighbours found can be a lower limit, but we assume that the missing neighbours without redshifts are evenly distributed over our LSBGs sample, which seems reasonable for our statistical analysis. The results are listed in Table~\ref{tab:sample}, in the column labeled ``Iso", for isolation parameter. 

\begin{figure*} 
\centering
\includegraphics[scale=0.58]{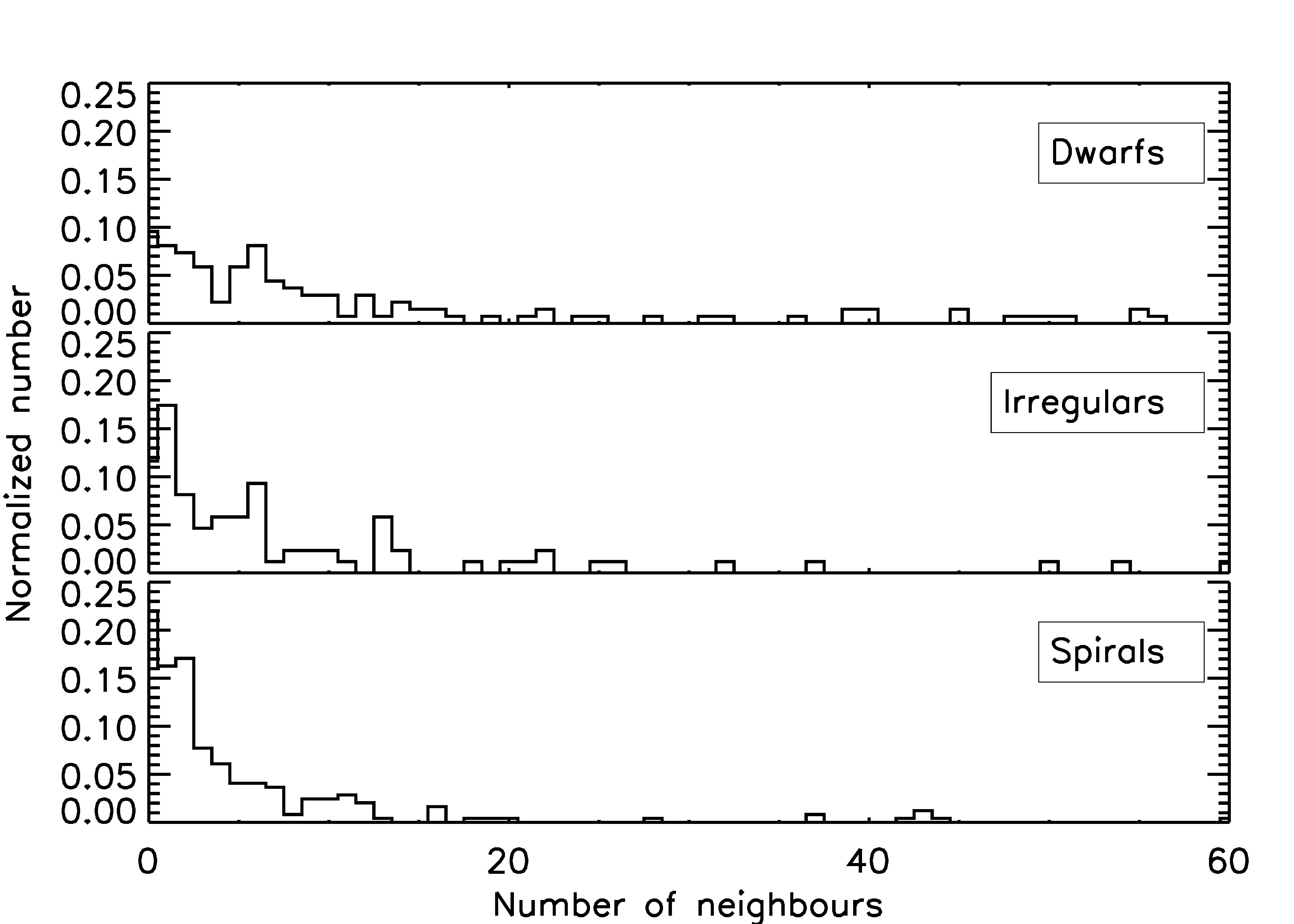}\includegraphics[scale=0.51]{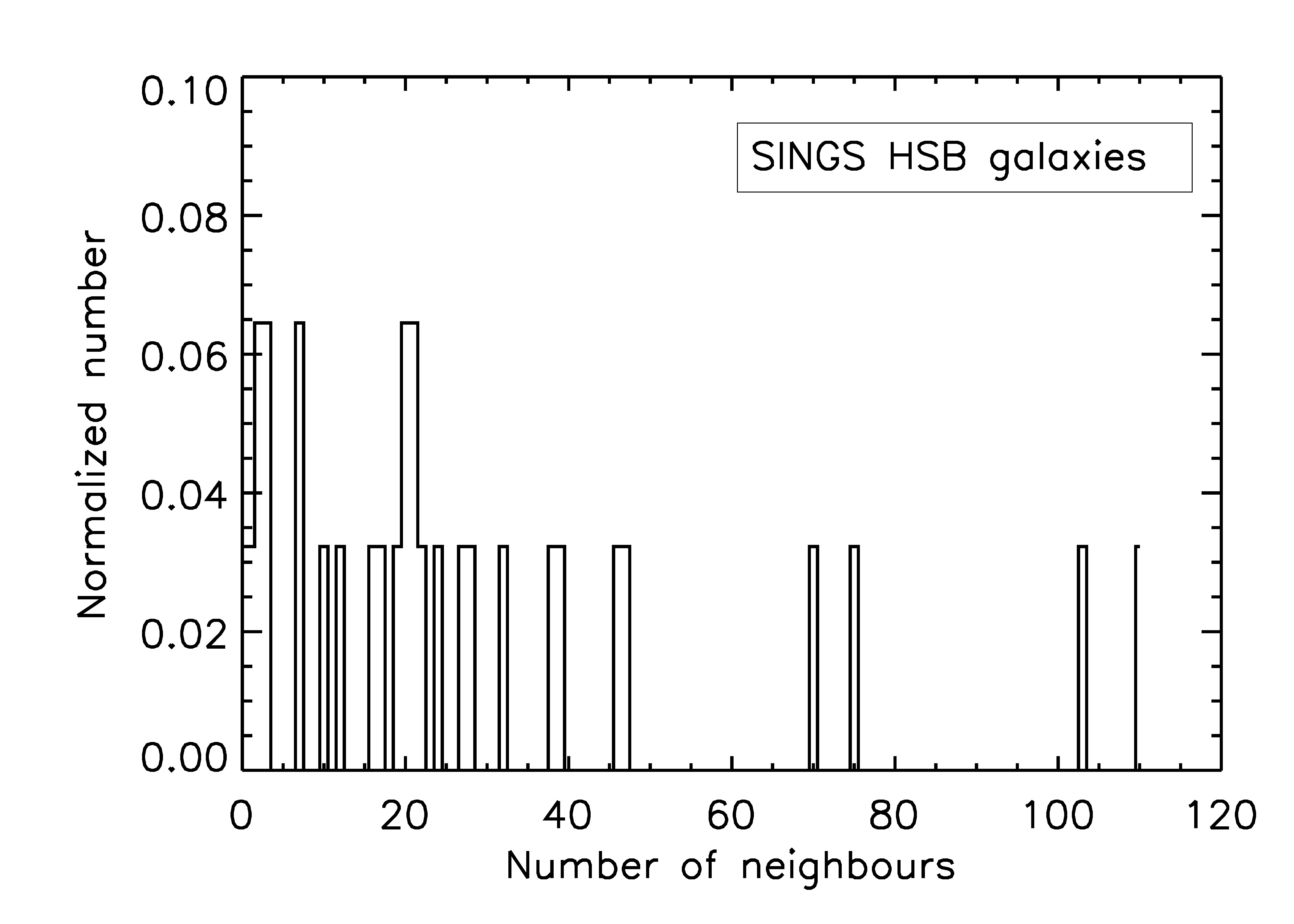}
\caption{Histograms of the normalized distribution of the number of nearby neighbours of sample galaxies.  Nearby neighbours are all known galaxies within a radius of 1 Mpc and within $\pm$500 \kms\ of the systemic velocity of the sample galaxy. In each histogram, the distribution is normalized by the total number of galaxies in the  histogram. Shown are data for LSBGs of three different morphological classes (left panel), and for HSB galaxies from the SINGS sample (right panel). We used 246 LSBG spirals, 86 irregulars and 136 dwarfs for this plot, and 37 HSB galaxies.}    
\label{fig:iso_histogram}
\end{figure*}

As we have used straightforward Hubble flow distances for all our previous calculations and plots, we also used these as input values for the NED search routine. However, in this case the routine limits the distance range for target galaxies from 10 to 150 Mpc. For this reason, the sample of galaxies for which we could determine the number of near neighbours is reduced to 468. It includes 246 spirals, 86 irregulars and 136 dwarfs. Figure~\ref{fig:iso_histogram} shows histograms of the number of near neighbours normalized by the total number of nearby galaxies found, for spirals, dwarfs and irregulars.

We find that 63\% of LSB spirals have $\le$3 neighbours and 87\% have $\le$10 neighbours, whereas for the irregulars and dwarfs the corresponding numbers are 42\% and 71\%, and 31\% and 61\%, respectively. Thus, LSB spirals are the most isolated class and dwarfs are the least isolated of LSBGs. 

We compared the isolation of galaxies in our LSBG sample with that of normal galaxies using the Spitzer Infrared Nearby Galaxies Survey (SINGS,  \citealt{kennicutt.etal.2003}; see Fig.~\ref{fig:iso_histogram}). We used 37 galaxies from this survey (at distances between 10 and 150 Mpc) and estimated the number of their near neighbours in the same way as for our sample. Only 19\% of the SINGS galaxies have $\le$3 neighbours and 29\% have $\le$10. 
Thus it is clear that our LSBGs have fewer neighbours than HSB galaxies, or in other words are more isolated than  normal galaxies. This is in agreement with previous studies \citep{Schombert.etal.1988, Schombert.etal.1992, Bothun.etal.1993, Rosenbaum.etal.2009, Galaz.2011}. 
The isolation of LSBGs is one of the main reasons why they have remained so poorly evolved, since they do not have interactions with other galaxies \citep{Zaritsky1993}. Their isolated nature and high dark matter content results in a lower rate  of disc instability formation, both local and global \citep{wadsley.mayer.2004, ghosh.jog.2014}, which means that the disc SFR is lower, thereby causing disc evolution to be slower \citep{Galaz.2011}. 

\subsubsection{Gas mass fraction as a function of environment} \label{subsubsec:gasfraction_environment} 
\begin{figure} 
\centering
\includegraphics[scale=0.5]{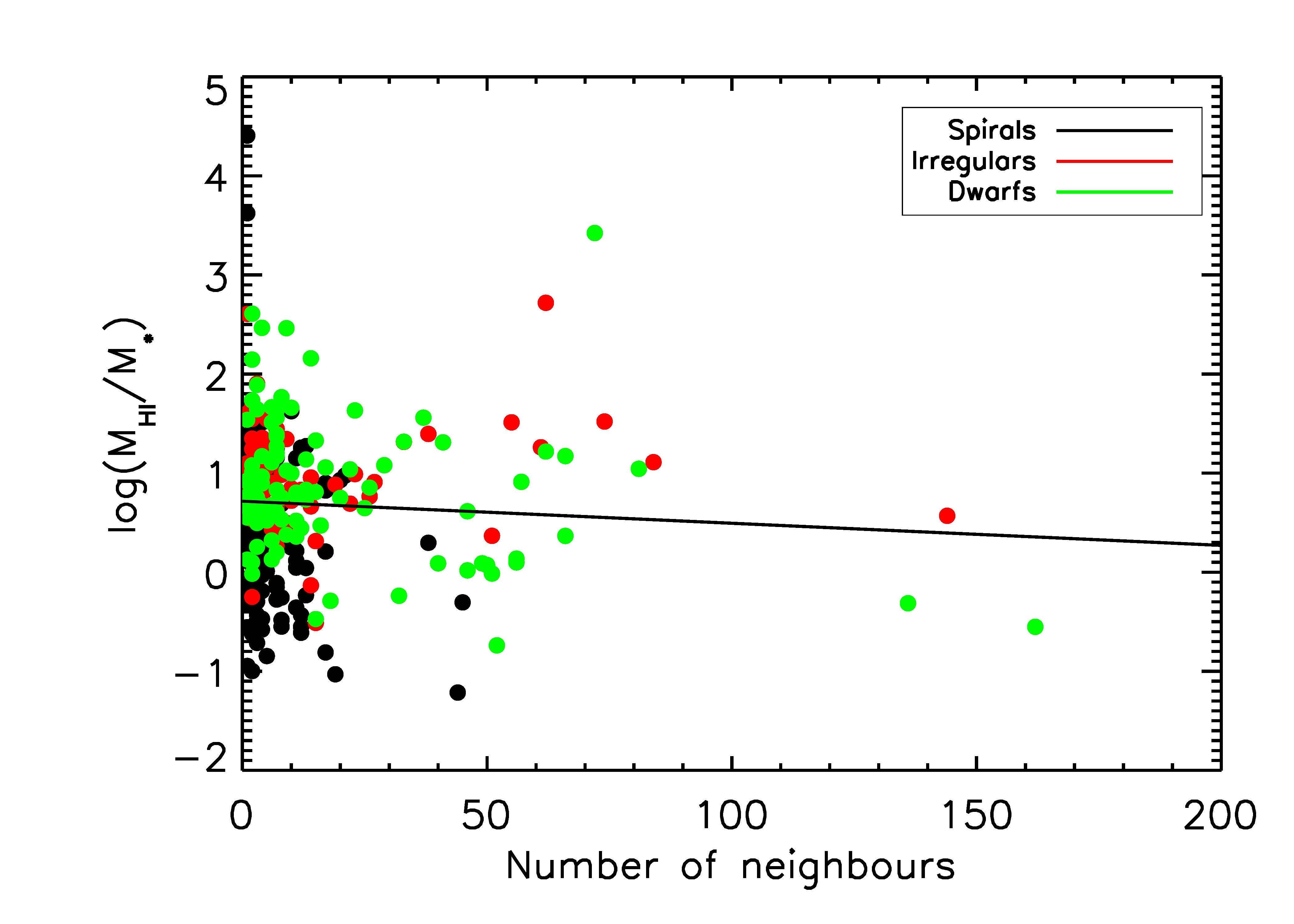} \includegraphics[scale=0.5]{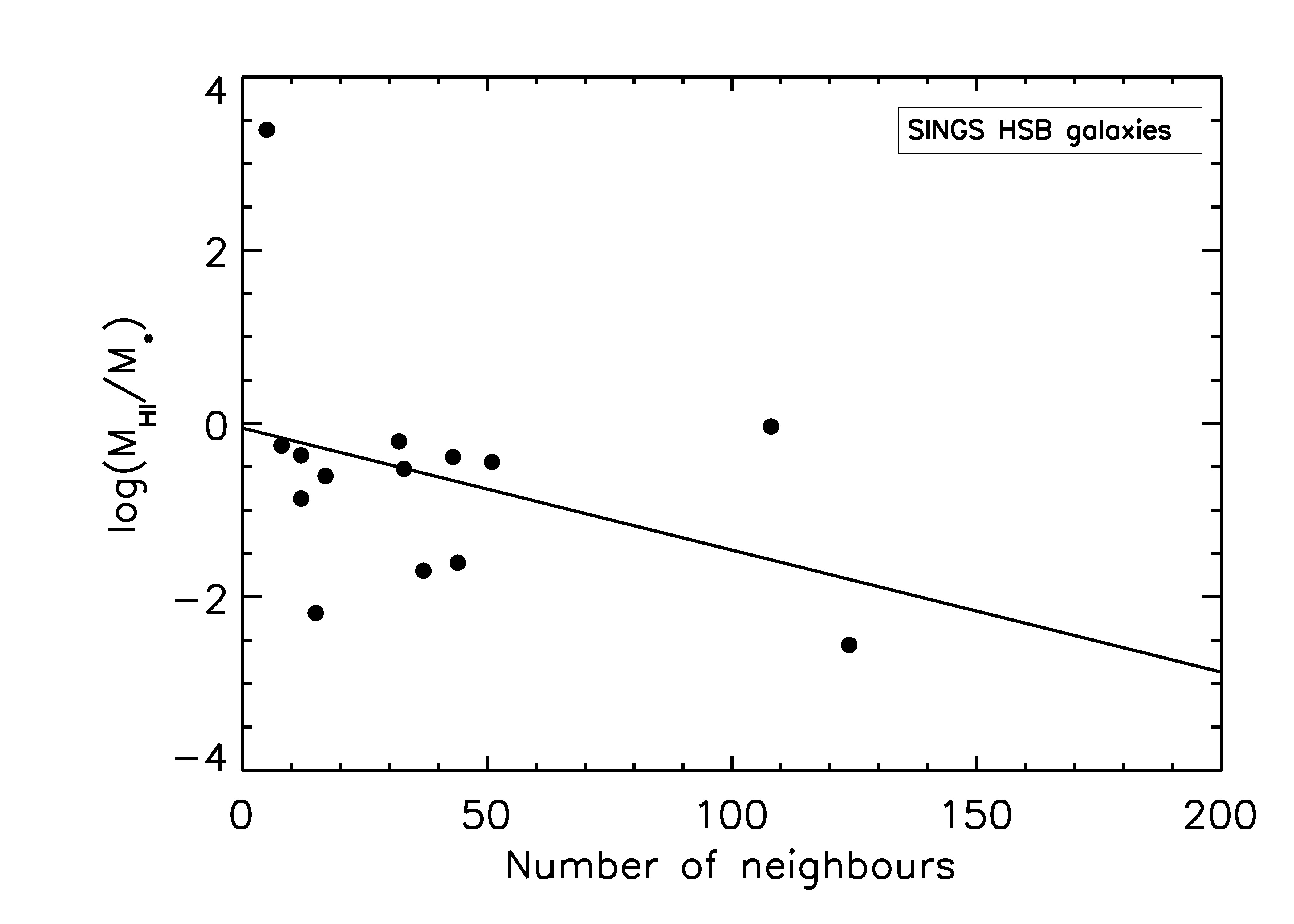}
\caption{Gas fraction, log(\MHI/\Mstar), as a function of the number of near neighbours, for our LSBGs (top panel) and for SINGS HSB galaxies (bottom panel). We used 163 LSBGs spirals, 71 irregulars and 102 dwarfs for this plot, and 37 HSB galaxies. The linear LSQ fit to the LSBGs data shows a nearly flat slope $-$0.0022$\pm$0.0020 and the fit for the SINGS galaxies a steeper slope of $-$0.014$\pm$0.010.}
\label{fig:HI_fraction_iso}
\end{figure}

To examine whether interactions are important for triggering star formation in LSBGs we plotted the HI mass fraction against the number of nearby neighbours ($<$~1~Mpc), see Figure~\ref{fig:HI_fraction_iso}. There is a lot of scatter and there does not appear to be much dependence of \MHI/\Mstar\ on the number of near neighbours. A line fitted to this distribution is almost flat, with a slope of $-$0.0022$\pm$0.0020. 
This suggests that the environment does not have a pronounced effect on converting HI mass into stellar mass in LSBGs. We found that SINGS galaxies exhibit a slightly steeper slope of $-$0.014$\pm$0.010. 

The fact that LSBGs have extended HI discs indicates that they are isolated and unperturbed systems.  For example,  HI discs in LSB spirals are at least two times larger than their optical radii \citep{pickering.etal.1997, auld.etal.2006, das.etal.2007, mishra.etal.2017} and in dwarfs they are even more extended  \citep{Begum2008}. 

\subsection{Giant LSB galaxies} \label{subsec:glsb}
Using  selection criteria of log(\MHI) $>$ 9 and log(\Mstar) $>$ 10 to define giant LSB galaxies (see e.g., \citealt{matthews.etal.2001}; \citealt{das.2013}) we find that 41 out of 426 (or 9.6\%) of our LSBGs are giants; we will refer to them as GLSBGs. 

In Table~\ref{tab:GLSB} we list the following global properties of the 41 GLSB galaxies identified in our sample, see Sect. \ref{sec:analysis} for further details: 

\begin{itemize}
\item{Galaxy name}; 
\item{\Mstar}: total stellar mass (in \Msun); 
\item{\MHI}: total HI mass (in \Msun); 
\item{$Iso$}: isolation parameter (see Sect.~ \ref{subsec:Environment}), determined for galaxies with distances between 10 and 150 Mpc;
\item{$Iso$$^*$}: isolation parameter determined for galaxies at distances $>$150 Mpc, using the NED methodology for determining distances, which is based on the same Hubble constant used throughout this paper but with added standard cosmological parameters; 
\item{$D_{\rm K_{\rm s}}$}: major axis diameter in the $K_{\rm s}$ band, from  2MASS, the Two Micron All Sky Survey  (in kpc); in case only $R$-band values were available these are indicated by ($R$);
\item{$D_{\rm 25}$}: major axis diameter at the 25.0 magnitude arcsec$^{-2}$ isophote level on blue POSS-I images, as listed in the UGC (\citealt{Nilson.1973}, Uppsala General Catalogue of Galaxies (UGC)), in kpc. 
If not available, we listed  the $B$-band values given in the Third Reference Catalogue of Galaxies (\citealt{deVaucouleurs.1991}, RC3), indicated by ($B$).
\end{itemize}

The mean HI mass of the GLSBGs in our sample is log(\MHI) = 10.14$\pm$0.53. This is about 2.3 times larger than the mean of all our spirals, and 8.7 and 15.5 times that of the irregulars and dwarfs, respectively.
The mean stellar mass of the GLSBGs is log(\Mstar) = 10.60$\pm$0.41, or about 5.4 times the mean value for our spirals and about 102 and 135 times that for our irregulars and dwarfs, respectively (see Sect.~\ref{subsec:stellarmass}).

Giant LSB galaxies are usually redder than  normal sized LSB galaxies and appear to have an evolved stellar population similar to HSB galaxies \citep{Sprayberry.etal.1995, vandenhoek.etal.2000}. 
They are unique not just because of their large mass and size, but also because of their high dark matter content.

 \begin{table}  
{\tiny
 \caption{Global properties of the giant LSB galaxies.} 
\label{tab:GLSB}
 \begin{tabular}{lcccccc}
\hline 
galaxy name & log(\Mstar) & log(\MHI) & $Iso$ & $Iso^*$ & $D_{\rm K_{\rm s}}$ & $D_{\rm 25}$ \\
& (\Msun) & (\Msun) & & & (kpc) & (kpc) \\
\hline 
LSBC F611-03 & 10.12 & 10.15 & ...&  4   &  28.6      &  ...	  \\
LSBC F612-V02& 10.78 & 10.36 & ...&  2   &  66.5      &  ...	  \\
0223-0033    & 10.90 & 10.05 & 5  &      &  47.4      &  64.9	  \\
LSBC F560-05 & 11.29 & 10.11 & ...&  2   &  38.7      &  ...	  \\ 
LSBC F560-01 & 10.11 & 9.97  &1   &      &  38.9      &  28.8	  \\
IC 2409      & 10.21 & 10.02 & 4  &      &  36.3      &  25.6	  \\
IC 2423      & 10.52 & 9.97  &8   &      &  43.2      &  40.9	  \\
LSBC F564-02 & 10.16 & 10.41 & ...&  4   &  74.5($R$) &  ...    \\	    
IC 2454      & 10.39 & 9.51  &38  &      &  36.2      &  42.6	  \\         
UGC 5035     & 10.70 & 10.00 & ...&  6   &  60.2      &  ...	    \\
1003+0151    & 10.51 & 11.03 & ...&  1   &  48.6($R$) &  ...	  \\
LSBC F638-02 & 10.15 & 9.59  &1   &      &  25.0      &  ...	  \\  
LSBC F568-09 & 10.23 & 9.70  &5   &      &  ...       &  60.9	  \\    
LSBC F568-08 & 10.72 & 10.23 & 3  &      &  60.2      &  42.3	  \\
LSBC F568-06 & 10.88 & 10.57 & ...&  3   &  66.8      &  ...	  \\
NGC 3639     & 10.13 & 9.97  &3   &      &  37.4      &  20.3	  \\
UGC 6614     & 10.38 & 10.15 & 13 &      &  23.7      &  87.9	  \\
NGC 3821     & 10.80 & 9.13  &44  &      &  36.8      &  37.7   \\  
NGC 3939     & 10.37 & 9.42  &1   &      &  31.7      &  28.3	  \\         
IC 742       & 10.45 & 9.23  &44  &      &  39.0      &  36.1	  \\  
1213+0127    & 10.32 & 9.76  & ...&  3   &  39.5      &  ...	  \\ 
LSBC F574-05 & 10.77 & 10.03 & ...&  2   &  48.7      &  ...	  \\ 
1252+0230    & 10.31 & 10.20 & ...&  6   &  36.1      &  33.2	  \\
1300+0055    & 10.27 & 9.49  & ...&  2   &  49.9      &  ...	  \\
1300+0144    & 10.29 & 10.05 & ...&  3   &  44.2      &  ...	  \\ 
UGC 8794     & 10.43 & 9.95  &8   &      &  35.3      &  61.1	  \\
UGC 8828     & 10.72 & 9.91  &17  &      &  70.7      &  57.8($B$)\\        
LSBC F579-01 & 10.30 & 9.87  &3   &      &  39.2      &  ...	  \\
LSBC F579-03 & 10.38 & 9.74  &5   &      &  31.9      &  36.8	  \\ 
IC 1021      & 10.65 & 9.65  &2   &      &  49.7      &  38.3    \\  
LSBC F579-V01& 11.25 & 9.36  &3   &      &  20.6($R$) &  ...	  \\
UGC 9503     & 10.28 & 9.73  &12  &      &  33.1      &  60.4($B$) \\   
UGC 9634     & 10.59 & 10.49 & ...&  2   &  44.3      &  54.4	  \\
UGC 9843     & 10.29 & 10.15 & ...&  26  &  39.4      &  44.4($B$) \\
LSBC F584-01 & 10.62 & 9.90  & ...&  7   &  36.1      &  40.1($B$) \\  
LSBC F727-V04& 10.69 & 10.20 & ...&  7   &  44.5      &  ...	  \\
UGC 10405    & 10.47 & 10.49 & ...&  3   &  ...       &  82.0	  \\
LSBC F533-03 & 10.43 & 10.30 & ...&  1   &  82.2      &  47.0($B$) \\
LSBC F746-02 & 10.54 & 10.06 & ...&  1   &  28.0      &  ...	  \\  
LSBC F675-01 & 10.00 & 9.96  & ...&  2   &  26.0      &  ...	  \\ 
2327-0244    & 10.72 & 10.10 & 2  &      &  63.9      &  75.2($B$) \\    
\hline 
\end{tabular}
}
\end{table}

They are generally isolated \citep{Rosenbaum.etal.2009}. To see if our 41 GLSB spirals are more isolated than the other LSBGs in our sample, we examined their number of nearby neighbours (see Sect.~\ref{subsec:Environment}). We find that 73\% (30 out of 41) GLSBs have $\le$5 neighbours, whereas it is 58\% for the overall spiral LSBG sample. Thus GLSB galaxies represent the most isolated subsample of LSBGs. These giant field galaxies evolve mainly due to internal processes, since external interactions will be very rare. Hence they are the best systems in which to study the slow, secular evolution of disc galaxies, especially those with bars \citep{honey.etal.2016}. Their evolution may thus be different from the other LSB galaxies, especially the dwarfs and irregulars.
 
\subsection{Are LSB spirals distinct from dwarfs and irregulars? A statistical test} \label{subsec:kstest} 

From our results presented in the earlier sections it is clear that LSB spirals are distinct in many ways from the LSB dwarfs and irregulars -- for example, their mean stellar mass is about 20 times larger. To further quantify this difference we performed  Kolmogorov-Smirnov (KS) statistical tests to determine whether the LSB subclasses are all derived from the same parent sample \citep{wall.1996}. We performed the two-sample KS tests using the {\sc idl} code {\sc kstwo.pro}\footnote{https://idlastro.gsfc.nasa.gov/ftp/pro/math/kstwo.pro}, on the HI mass, stellar mass and  isolation parameter in pair-wise comparisons and summarized our results in Table~\ref{tab:kstest}. We found that based on the HI and stellar masses, it is virtually impossible (99.9\%) that LSB spirals arise from the same galaxy distribution as dwarfs and irregulars. The isolation parameter gives a very small probability (0.5\%) that they are from the same distribution. However, the LSB dwarfs and irregulars have a significant probability (up to 18\%) that they belong to the same distribution for all three parameters that we examined. 

\subsection{Implications of the statistical  test for the origin of LSBGs} \label{subsec:implications}  
One the reasons for performing the KS test on our sample is to explore the origin of the class of LSB galaxies. To date, the following hypotheses have been put forward regarding their formation: (i)~major merger of two face-on disc galaxies \citep{mapelli.etal.2008}, (ii)~the accretion of small galaxies by a large LSBG \citep{kaviraj.etal.2014}, (iii)~the evolution of rare 3$\sigma$ peaks in isolated regions or voids in the primordial density field of large scale structures  \citep{hoffman.silk.1992} and (iv)~they are disc galaxies that formed in fast rotating halos with high specific angular momentum,  resulting in low surface density discs, a common characteristic of LSBGs \citep{kim.lee.2013}. 

However, both major and minor mergers (scenarios i and ii) will destroy or distort the galaxy discs and lead to major star formation \citep{Rodriguez-Gomez.etal.2016}, thus transforming the LSB nature of these galaxies into HSB. Hence, scenarios (iii) and (iv) are more likely explanations for the origin of LSBGs. Of these two, the isolated 3$\sigma$ peaks in the primordial large scale structures'  density field could have resulted in the formation of large, or even giant, LSBGs that are extremely isolated. However, both the smaller and larger LSBGs may also have formed as discs in fast rotating halos; this origin scenario has no dependence on environment. Thus it appears that only the more massive, spiral LSBGs can form in isolated environments. This matches the results of our KS tests, which indicates that LSB spirals as a galaxy population are different from dwarfs and irregulars.

\begin{table}  
 \caption{Results of Kolmogorov-Smirnov tests on the probability that different morphological types of LSBGs originated  from the  same distribution.}
\label{tab:kstest}
 \begin{tabular}{lccc}
\hline
\multirow{2}{*}{Pair }&\multicolumn{3}{c}{probability of having same distribution in} \\
 		 & HI mass  & stellar mass  &   $Iso$  \\
\hline 

spirals \& irregulars & 3 $10^{-15}$	 & 6 $10^{-23}$	 & 5 $10^{-3}$	 \\

spirals \& dwarfs     & 2 $10^{-39}$ 	 & 4 $10^{-40}$ & 1 $10^{-10}$ \\
irregulars \& dwarfs  & 3 $10^{-3}$	   &  0.18	       & 0.06	\\
\hline
\end{tabular}
\end{table}

Although LSB dwarfs and irregulars have been found in both cluster and field environments \citep{Pustilink.etal.2011, Merritt.etal.2014, giallongo.etal.2015, davies.etal.2016, Wittmann.etal.2017}, in order  to retain their LSB nature the galaxies must remain fairly isolated. Recently a population of  ultra diffuse galaxies (UDGs) with stellar masses like dwarfs but radii like $L{\star}$ galaxies has been identified in nearby galaxy clusters and groups (e.g., \citealt{koda.etal.2015}; \citealt{trujillo2017}), some of which may be dwarf or irregular LSBGs \citep{Amorisco.Loeb.2016}. On the other hand,  many dwarf LSBGs have also been detected in voids \citep{kreckel.etal.2011}. The larger LSBGs are however nearly always isolated \citep{pickering.etal.1997}.

Regarding the considerably more massive LSB spirals, they may well be disc galaxies of masses similar to those of HSB spirals that are relatively isolated and therefore free from external shocks, and thus evolving in a more quiescent manner into red and gas-rich systems.

\section{Conclusions} 	\label{sec:conclusion}
The morphologies of LSBGs can be broadly classified as spirals, dwarfs and irregulars.
In terms of both total HI and stellar mass, we find that the LSB spirals are the most massive, followed by the irregulars, and then the dwarfs. 

The $r$-band luminosity increases with HI mass in all three morphological types. 
When taken as a whole, we find that the spirals show a different slope compared to irregulars and dwarfs in the \MHI/\Mstar--\Mstar\ plane. Within the total stellar mass estimation uncertainties, the fit parameters for our LSBGs are similar to the samples of \citet{Papastergis.etal.2012} and \citet{van2016}, which are dominated by HSB spirals.

The integrated ($g$--$r$) galaxy colours do not show a clear dependence on HI mass in LSB spirals, dwarfs and irregulars. Only the most massive spirals appear slightly redder, which indicates that even when they have a massive HI content, LSBGs cannot form stars at a significant rate. This can be due to the low surface density of their stellar and HI discs, as well as the presence of massive dark matter halos. 

The colour magnitude diagram (CMD) for LSBGs shows that spirals occur in both the blue cloud and the red sequence, indicating the variety in their evolutionary stages, 
with the red spirals being more massive (both  \Mstar\ and \MHI\  $\ge$ 10$^{9.5} $\Msun) and the blue ones less so (\Mstar\ $\sim$ 10$^7$-10$^9$ \Msun, \MHI\ $\approxgt$ 10$^8$ \Msun).
A significant fraction of the LSB spirals are very red, similar to galaxies in which star formation is quenched. The smaller LSB dwarfs and irregulars are comparatively bluer, indicating that they support slow and continuous star formation.

Exploring the nearby galaxy density around LSBGs, within 1~Mpc radius and $\pm$500 \kms\ of their systemic velocity, we confirmed that LSBGs are more isolated than HSB galaxies. We also found that the LSB spirals are more isolated than dwarf and irregulars, and that giant LSBGs are the most isolated of all.

Kolmogorov-Smirnov (KS) tests using HI masses, stellar masses and isolation parameters indicate that that the LSB spirals form a distinct population from the dwarfs and irregulars. This suggests a different formation process and/or evolutionary path for spirals compared to dwarfs and irregulars.

\vspace{3mm}
 
{\bf \sc Acknowledgments}
We wish to thank the referee Greg Bothun for his valuable comments and suggestions. We thank our colleague Vaibhav Pant for helpful discussions. We have used the NASA/IPAC Extragalactic Database (NED), which is operated by the Jet Propulsion Laboratory, California Institute of Technology, under contract with the National Aeronautics and Space Administration, as well as the online HyperLeda database. This work has used SDSS-III data. Funding for SDSS-III has been provided by the Alfred P. Sloan Foundation, the Participating Institutions, the National Science Foundation, and the U.S. Department of Energy Office of Science. 


\newpage
\onecolumn

\begin{landscape}

\begin{longtable}{lcrlccccrrrl}
\setcounter{table}{1}
\label{tab:samplefull}
\endfirsthead
\caption{Global properties of the LSBGs in the initial sample of 897 galaxies-- {\it full version}.}\\
\hline 
galaxy name & RA \& DEC & $D$ & morphology & class & HI flux & log(\MHI)& \Mr\ & ($g$--$r$) & log(\Mstar)& $Iso$ & Flag$^*$ \\

 & (J2000.0) & (Mpc) & $[$NED$]$ &  & (Jy km/s) & (\Msun) & (mag) & (mag) & (\Msun) & & \\ 
\hline 
\endfirsthead
\caption{-- {\it continued}.}\\
\hline 
galaxy name & RA \& DEC & $D$ & morphology & class & HI flux & log(\MHI)& \Mr\ & ($g$--$r$) & log(\Mstar)& $Iso$ & Flag$^*$ \\
 & (J2000.0) & (Mpc) & $[$NED$]$ &  & (Jy km/s) & (\Msun) & (mag) & (mag) & (\Msun) & & \\ 
\hline 
\endhead
\hline
\endfoot
\multicolumn{12}{l}{1. With HI detections:}	 \\ 
LSBC F539-V01 	&00:11:29.00 +21:25:52.0 & 113.3 & Sm & spiral &2.37 & 9.86 & $-$18.52 & 0.34 &8.75 & 	1 & good \\
LSBC F608-V01 	&00:12:48.25 +14:31:31.1 & 26.1 & Sd & dwarf &3.47 & 8.75 & $-$14.71 & 0.24 &7.09 & 	10 & good \\
0012+0218 	&00:14:59.97 +02:34:48.3 & 256.3 & dI & dwarf &0.69 & 10.03 & $-$19.15 & 0.36 &9.02 & ... & good \\
0014+0210 	&00:16:41.72 +02:27:34.4 & 59.1 & Sm: & star	 &0.75 & 8.79 & $-$15.62 & 1.34 &8.91 & 	1 & not used\\	
LSBC F608-01 	&00:17:15.90 +17:31:23.0 & 14.3 & Im: & irregular &2.18 & 8.02 & $-$13.99 & 0.92 &7.71 & 	15 & not used \\
LSBC F473-V01 	&00:17:18.30 +26:51:41.0 & 127.3 & Im & irregular &1.36 & 9.72 & $-$18.57 & 0.33 &8.76 & 	1 & good \\
LSBC F608-V02 	&00:23:06.06 +15:08:25.6 & 78.8 & Sd & irregular &0.85 & 9.09 & $-$17.05 & 0.50 &8.37 & 	10 & good \\
\end{longtable}
\end{landscape}

\end{document}